\documentclass[aps,10pt,twocolumn,superscriptaddress, nofootinbib]{revtex4-2}
\usepackage{mathrsfs, amssymb, amsmath, mathtools}  
\usepackage{epsfig, cancel, comment}
\usepackage{footmisc}
\usepackage{latexsym}
\usepackage{natbib}
\usepackage{url}
\usepackage{dcolumn}
\usepackage{multirow}
\usepackage{color}
\usepackage{cancel}
\usepackage{soul}
\usepackage[normalem]{ulem}
\usepackage{amsfonts,amssymb,amsmath, txfonts}
\usepackage{graphicx,epsfig}
\usepackage{psfrag}
\usepackage{hyperref}
\usepackage{mathtools}
\usepackage{enumitem}
\usepackage{float}
\usepackage[dvipsnames]{xcolor}
\usepackage{xcolor}
\hypersetup{linktoc=all,
    colorlinks=true, linkcolor={brightpink},
    citecolor={blue}, urlcolor={blue}
}
\definecolor{rosy}{RGB}{230,235,252}
\definecolor{myframetitle}{RGB}{90,89,170}
\definecolor{myblocktitle}{RGB}{140,185,249}
\definecolor{mytitle}{RGB}{10,80,26}

\definecolor{darkgreen}{RGB}{27,130,45}
\definecolor{darkblue}{rgb}{0,0,0.3}
\definecolor{darkred}{rgb}{0.7,0,0}

\definecolor{light gray}{RGB}{220,220,220}
\definecolor{dark purple}{RGB}{108,0,217}
\definecolor{pink}{RGB}{190,20,100}
\definecolor{orang}{RGB}{193,63,0}
\definecolor{green}{RGB}{11,98,17}
\definecolor{darkpink}{RGB}{153,0,76}
\definecolor{bluegreen}{RGB}{0,102,102}
\definecolor{greenlagan}{RGB}{0,102,0}
\definecolor{redgreen}{RGB}{102,102,0}
\definecolor{Redgreen}{RGB}{153,76,0}
\definecolor{vividviolet}{rgb}{0.62, 0.0, 1.0}
\definecolor{amaranth}{rgb}{0.9, 0.17, 0.31}
\definecolor{palatinateblue}{rgb}{0.15, 0.23, 0.89}
\definecolor{brightpink}{rgb}{1.0, 0.0, 0.5}
\definecolor{cornflowerblue}{rgb}{0.39, 0.58, 0.93}
\definecolor{deepcarminepink}{rgb}{0.94, 0.19, 0.22}
\definecolor{radicalred}{rgb}{1.0, 0.21, 0.37}

\def\H0{{\text{H}\hspace*{-2.05mm}\text{H} 0\hspace*{-1.35mm}0\ }}

\def\be{\begin{equation}}
\def\ee{\end{equation}}
\def\beq{\begin{equation}}
\def\eeq{\end{equation}}
\def\bea{\begin{eqnarray}}
\def\eea{\end{eqnarray}}
\newcommand{\dd}{\textrm{d}}

\begin{document}

\title{$\Lambda$CDM Tensions: Localising Missing Physics through Consistency Checks}

\author{\"{O}. Akarsu}
\affiliation{$^{2}$Department of Physics, Istanbul Technical University, Maslak 34469 Istanbul, Turkey}
\author{E. \'O Colg\'ain}
\affiliation{Atlantic Technological University, Ash Lane, Sligo F91 YW50, Ireland}
\author{A. A. Sen}
\affiliation{Centre for Theoretical Physics, Jamia Millia Islamia, New Delhi - 110025, India}
\author{M. M. Sheikh-Jabbari}
\affiliation{School of Physics, Institute for Research in Fundamental Sciences (IPM),\\ P.O.Box 19395-5531, Tehran, Iran}

\begin{abstract}
$\Lambda$CDM tensions are by definition model dependent; one sees anomalies through the prism of $\Lambda$CDM. Thus, progress towards tension resolution necessitates checking the consistency of the $\Lambda$CDM model to localise missing physics either in redshift or scale. Since the Universe is dynamical and redshift is a proxy for time, it is imperative to first perform consistency checks involving redshift, then consistency checks involving scale, as the next steps to settle the ``systematics versus new physics" debate and foster informed model building. We present a review of the hierarchy of assumptions underlying the $\Lambda$CDM cosmological model and comment on whether relaxing them can address the tensions. We focus on the lowest lying fruit of identifying missing physics through the identification of redshift dependent $\Lambda$CDM model fitting parameters. We highlight recent progress made on ${S_8:= \sigma_8 \sqrt{\Omega_{\rm m}/0.3}}$ tension and elucidate how similar progress can be made on $H_0$ tension. Our discussions indicate that $H_0$ tension, equivalently a redshift dependent $H_0$, and a redshift dependent $S_8$ imply a problem with background $\Lambda$CDM cosmology. 
\end{abstract}

\maketitle

\begin{raggedright} {\textit{``Will you walk into my parlour?" said a spider to a fly;\\
``'Tis the prettiest little parlour that ever you did spy.\\
The way into my parlour is up a winding stair,\\
And I have many pretty things to show when you are there."\\ 
``Oh no, no!" said the little fly, ``to ask me is in vain,\\
For who goes up your winding stair can ne'er come down again."}-- The Spider and the Fly, Mary Howitt (1829) }\end{raggedright}

\section{Introduction}
In the absence of firm theoretical guidance, cosmology makes progress through Occam's razor and Bayesian model comparison~\cite{Trotta:2005ar, Hobson:2009, Trotta:2017wnx} \footnote{See \cite{Amendola:2024prl} for a sharp discussion on the arbitrariness inherent to ratios of Bayesian evidences. In particular, when there is poor theoretical guidance on parameter space, it seems possible to relax the priors so that the minimal model, e. g. $\Lambda$CDM, in any nested model is preferred \cite{Patel:2024odo}.}. To play this game, one identifies a cosmological model that has (i) a minimal degree of complexity, as quantified by counting free parameters, and (ii) a physics backstory and narrative. The standard model of cosmology today, the Lambda Cold Dark Matter ($\Lambda$CDM) model,   
represents the minimal model that explains Cosmic Microwave Background (CMB) data~\cite{Planck:2018vyg}. 
Both cold dark matter and late-time accelerated expansion driven by dark energy (DE) are observationally well-motivated in a spatially homogeneous and isotropic Universe, while the cosmological constant $\Lambda$ has theoretical pedigree, since it can be added to the Einstein field equations for free. Recently, in addition to the long-standing, notoriously challenging theoretical issues related to $\Lambda$~\cite{Weinberg:1988cp,Weinberg:2000yb,Peebles:2002gy}, observational discrepancies and tensions involving $\Lambda$CDM model parameters have emerged between data sets from different cosmic epochs~\cite{DiValentino:2021izs,Perivolaropoulos:2021jda, Abdalla:2022yfr}. If not due to systematics, the tensions point to missing physics not incorporated by the $\Lambda$CDM model.

The prevailing responses of the cosmology community to $\Lambda$CDM tensions have either involved studying systematics or the exploration of models with new fundamental physics~\cite{DiValentino:2021izs, Schoneberg:2021qvd}.  Nevertheless, less effort has been made to \textit{localise} or narrow down the effects of missing physics either in redshift, a proxy for cosmic time, or scale, a proxy for cosmic distances. On the contrary, one of the recognised front-running proposals to resolve the tension in the Hubble constant $H_0$~\cite{Planck:2018vyg, Riess:2021jrx, Freedman:2021ahq, Pesce:2020xfe, Kourkchi:2020iyz, Blakeslee:2021rqi} that invoke new physics pre-recombination~\cite{Knox:2019rjx, Poulin:2018cxd, Agrawal:2019lmo, Lin:2019qug, Niedermann:2019olb, Ye:2020btb, Poulin:2023lkg} effectively \textit{delocalise} missing physics in epochs with at best indirect observational constraints. One turns the $H_0$ tension from an apparent problem in the redshift range $0 \lesssim z \lesssim 1100$ \footnote{Combining supernovae (SNe) and BAO constraints one may be able to whittle this down to $1 \lesssim z \lesssim 1100$.} to a problem in the redshift range $1100 \lesssim z < \infty$, more generally $0 \lesssim z < \infty$; see \cite{DiValentino:2021izs,Perivolaropoulos:2021jda, Abdalla:2022yfr} for a comprehensive account of models injecting new physics at different epochs, and \cite{Vagnozzi:2023nrq} for arguments saying that one must introduce new physics in both the pre- and post-recombination Universe. Despite delocalising physics, from the perspective of Bayesian model comparison, these proposals are par for the course. They introduce a finite number of additional degrees of freedom, thereby making the proposals both competitive and testable, while also providing a requisite physics backstory. This modelling game is without end.  

Against this backdrop, this letter has two objectives. The first is to provide a review of the assumptions underlying $\Lambda$CDM cosmology and comment on whether relaxing them promises to alleviate $\Lambda$CDM tensions. The second objective is to highlight the low lying fruit of better localising missing physics within the $\Lambda$CDM model. This entails conducting consistency checks of the $\Lambda$CDM model at different redshifts (or scales),\footnote{In cosmology, redshift is more fundamental than scale, so one must eliminate redshift first, before turning one's attention to scale. For this reason, we focus largely on redshift.} where one looks for  signatures that the fitting parameters of the model are changing; redshift or scale dependence of the $\Lambda$CDM cosmological parameters signal an inconsistency and breakdown of the model. If signatures of drift or evolution of the fitting parameters can be found, this localises missing physics to the redshift ranges under study. If they cannot be found, the $\Lambda$CDM model performs well in the range. Furthermore, if one succeeds in localising missing physics to a set of measure zero, this swings the tensions debate back to systematics.  

For cosmologists familiar with Bayesian model comparison, the proposed checks of the $\Lambda$CDM model may sound unfamiliar. Admittedly, there are key differences. Notably, there is no second model and one is performing a consistency check of a single model to show that the model fitted to data at different redshift ranges (or scales) leads to consistent fitting parameters. Intuitively, if different subsamples of the data prefer different fitting parameters, this worsens the goodness of fit, so the methodology complements the Bayesian approach. To fully appreciate this, note that if one fits the $\Lambda$CDM model to data in different redshift bins, and one finds the same best fit parameters in all bins, there is little hope of finding a model that beats $\Lambda$CDM in model comparison. Only when redshift evolution of $\Lambda$CDM fitting parameters is established  an alternative model becomes
competitive. For this reason, it should be quicker to diagnose missing physics through redshift (or scale) dependent fitting parameters than construct a new model--necessitated by a Bayesian framework--with additional degrees of freedom that address the evolution. Bluntly put, in any race to diagnose $\Lambda$CDM tensions, Bayesian methods are expected to finish second best.             

\section{Hierarchy of Assumptions}
In this section, we provide a review of the assumptions that culminate in the $\Lambda$CDM cosmology. We comment on the prospects of relaxing these assumptions to resolve $H_0$ tension~\cite{Planck:2018vyg, Riess:2021jrx, Freedman:2021ahq, Pesce:2020xfe, Kourkchi:2020iyz, Blakeslee:2021rqi} and $S_8$ tension~\cite{Heymans:2013fya, Joudaki:2016mvz, DES:2017qwj, HSC:2018mrq, KiDS:2020suj, DES:2021wwk}. Our comments overlap with earlier observations in~\cite{Beenakker:2021vff}, but we expand them to include greater and more recent observational scope. 

\subsection{Cosmological Bedrock: GR + FLRW}

\textit{Could relaxing GR resolve $\Lambda$CDM tensions?} Modern cosmology rests upon General Relativity (GR) and the Einstein field equations, 
\begin{equation}
\label{eq:Einstein}
R_{\mu \nu} - \frac{1}{2} R g_{\mu \nu} + \Lambda g_{\mu \nu} = \frac{8 \pi G}{c^4} T_{\mu \nu},  
\end{equation}
where $g_{\mu \nu}$ is the spacetime metric, $R_{\mu \nu}$ and $R=g^{\mu\nu} R_{\mu \nu}$ are the Ricci (curvature) tensor and scalar, respectively, both derived from $g_{\mu \nu}$, $\Lambda$ is the cosmological constant, $T_{\mu \nu}$ is the total stress-energy tensor, $G$ is Newton's constant, and $c$ denotes the speed of light. $\Lambda$ is special because it may be consistently  added to the left hand side of the Einstein field equations.\footnote{The consistency condition of \eqref{eq:Einstein} is that both sides of this equation are divergence-free, i.e., $\nabla^{\mu} (R_{\mu \nu} - \frac{1}{2} R g_{\mu \nu})=0$ by virtue of Bianchi identity, and $\nabla^{\mu}(\Lambda g_{\mu \nu})=0$ as long as $\Lambda$ is a constant, implying that $\nabla_{\mu} T^{\mu \nu} =0$. Note overall consistency of \eqref{eq:Einstein} only implies the total stress-energy tensor to be divergence-free. Requiring individual components of the cosmic fluid, i.e., DE, pressureless matter and radiation, to have divergence-free stress tensors, is  an additional assumption.} As with any theory, one does not expect GR to hold universally. Notably, the formidable task remains to reconcile GR with quantum physics, the other key pillar of 20$^{\textrm{th}}$ century physics, and curvature singularities within black hole solutions point to regimes where the theory may break down. These observations have motivated extensive \textit{theoretical} speculation in the cosmology literature~\cite{Clifton:2011jh, Bamba:2012cp}. In contrast, \textit{observationally} GR has passed all tests with flying colours~\cite{Will:2014kxa}. GR is well tested to $\mu \rm m$ scales in the laboratory~\cite{Kapner:2006si, Lee:2020zjt}, in the weak-field setting in the solar system~\cite{Bertotti:2003rm}, the Milky Way~\cite{GRAVITY:2020gka}, and at extragalactic scales~\cite{Collett:2018gpf}. Moreover, GR has been tested in the strong-field regime with binary pulsars~\cite{Taylor:1979zz, Taylor:1982zz, Kramer:2021jcw} and gravitational waves~\cite{LIGOScientific:2019fpa, LIGOScientific:2020tif, LIGOScientific:2021sio}. It is imperative to continue to test the predictions of GR, but the absence of deviations makes the assumption (\ref{eq:Einstein}) remarkably solid \cite{Clifton:2005aj}. Despite investigations of the potential of modified gravity to resolve $\Lambda$CDM tensions~\cite{Rossi:2019lgt, Ballesteros:2020sik, Braglia:2020iik, Nguyen:2023fip}, the difficulties are well documented~\cite{Sakr:2021nja, Heisenberg:2022lob, Heisenberg:2022gqk, Lee:2022cyh}. 

Once one comes to terms with the fact that there is \textit{no} observational motivation \footnote{The slogan here is simple; theoretical yes, observational no!} for deviating from (\ref{eq:Einstein}), the next most fundamental assumption in modern cosmology is the \textit{cosmological principle}. Succinctly put, the Universe is a gravitating system, thus necessitating an assumption for $g_{\mu \nu}$. The consensus holds that, in accordance with the cosmological principle, at sufficiently large scales, the average evolution of the Universe is precisely described by the Friedmann-Lemaître-Robertson-Walker (FLRW) metric,
\begin{equation}
\label{eq:FLRW}
    g_{\mu \nu} \dd x^{\mu} \dd x^{\nu} = - c^2 \dd t^2 + a(t)^2 \dd \Sigma_{(3)}^2, 
\end{equation}
where {$t$ is the cosmic time}, $\Sigma_{(3)}$ is a maximally symmetric 3D space, and time dependence only enters through the scale factor $a=a(t)$. Note, $\Sigma_{(3)}$ need not be flat and it can have constant curvature, as hinted at by a spatially closed Universe in Planck data~\cite{Handley:2019tkm, DiValentino:2019qzk}. However, such an outcome drives $H_0$ to lower values, $H_0 \sim 50 {\rm \,km\, s^{-1}\, Mpc^{-1}}$, and sits uneasy with the inflationary paradigm, a key theoretical input in modern cosmology. In fact, inflation is not necessarily in conflict with a non-flat space, but spatially closed inflationary models are significantly fine-tuned~\cite{Linde:1995xm,Linde:2003hc}.\footnote{Modern cosmology blends data analysis with story telling in settings where observations do not exist. One could rewrite the inflationary backstory to allow for curved spacetimes, e.g.~\cite{Park:2017xbl}. We note that curvature can also be exploited to test FLRW~\cite{Clarkson:2007pz}.} For these reasons, we will focus on spatially flat $\Sigma_{(3)}$.\footnote{{Recently, the ACT collaboration analysing the CMB lensing data, resolved the so-called $A_{\textrm{L}}$ lensing anomaly of the Planck collaboration, providing further support for flat  cosmology~\cite{ACT:2023kun}. See also~\cite{Addison:2023fqc} for a recent investigation of the $A_{\textrm{L}}$ anomaly that unearths a connection to ecliptic latitude. This is curious because coincidences involving the ecliptic also appear in CMB anomalies~\cite{Schwarz:2015cma}, but the origin could easily be a systematic in Planck data.} } Once one inserts (\ref{eq:FLRW}) in (\ref{eq:Einstein}), $H_0$ emerges mathematically as an integration constant when one solves the Einstein $\rightarrow$ Friedmann equations~\cite{Krishnan:2020vaf}. Relaxing FLRW removes the requirement that the Universe's rate of expansion, as measured by the distance ladder through averages of the recession velocity over the sky, be a constant.\footnote{It is well known that in the simplest extensions of FLRW setting that allow for anisotropic expansion, the cosmic shear contributes as $a^{-6}$ in the Friedmann equations, and hence expansion anisotropies are expected to be theoretically more relevant in the early Universe, thereby delocalising physics and demanding a complete rewrite for the Universe's evolution \cite{Akarsu:2019pwn,Akarsu:2021max}. However, one should note that besides the shear, an anisotropy in the metric, an anisotropy can enter in the $T_{\mu\nu}$ part of cosmological models through the notion of the tilt \cite{King-Ellis}. A minimalistic, maximally Copernican beyond-FLRW model within the titled cosmology setting has been formulated as ``dipole cosmology'', which as the name suggests, accommodates cosmic dipoles in various matter sectors even in the late Universe \cite{Dipole-Cosmology, Dipole-Cosmology-2}. 
}


\textit{Could relaxing FLRW resolve $\Lambda$CDM tensions?} In principle, yes,\footnote{One can argue that resolutions to Hubble tension are strongly constrained by FLRW, the age of the Universe and the assumption of constant matter density~\cite{Krishnan:2021dyb}.} but this outcome would be most compelling if one succeeded in localising missing physics to the local Universe $z \lesssim 0.1$, because of the corroborating evidence for anisotropies in the same redshift range~\cite{Kashlinsky:2008ut, Migkas:2020fza, Migkas:2021zdo, Watkins:2023rll, Whitford:2023oww, Hoffman:2023pac}. Of course, if one relaxes the FLRW setup, the whole standard model of cosmology and its tensions should be redefined from scratch. That being said,  null tests of FLRW have been performed using radio galaxy and quasar (QSO) samples~\cite{Singal:2011dy, Rubart:2013tx, Singal:2019pqq, Secrest:2020has, Siewert:2020krp, Secrest:2022uvx}, arriving at findings at odds with FLRW, so one cannot preclude a violation of FLRW at distances beyond $500$ Mpc ($z \gtrsim 0.1$). In contrast to GR, where there are no anomalies worth reporting, recently some statistically significant deviations from FLRW behaviour have prompted debate~\cite{Kashlinsky:2008ut, Migkas:2020fza, Migkas:2021zdo, Watkins:2023rll, Whitford:2023oww, Hoffman:2023pac, Singal:2011dy, Rubart:2013tx, Singal:2019pqq, Secrest:2020has, Siewert:2020krp, Secrest:2022uvx, Pranav:2023khq, Mittal:2023xub}. Moreover, the independent anomalies appear synergistic and a preliminary science case has been built~\cite{Aluri:2022hzs}. Separately, leveraging longstanding CMB anomalies~\cite{Schwarz:2015cma}, it has been argued that a statistically anisotropic Universe cannot be dismissed~\cite{Jones:2023ncn}.\footnote{Note that the breakdown of cosmological principle (isotropy) can happen due to topology of constant time (spatial) slices while the metric is still FLRW, i.e., while the Universe is locally described by FLRW metric, e.g., see~\cite{COMPACT:2022gbl,COMPACT:2022nsu,COMPACT:2023rkp}. In particular, isotropy may be broken due to violation of parity as reported in~\cite{Jones:2023ncn}. Topology effects induce various relations among 2-point function correlations on the sky which are in principle detectable in the CMB or distributions of other large structures over the sky.} A plausible and conservative interpretation of some of these results is that FLRW is not a valid description of the local Universe out to at least 500 Mpc. This is enough to impact all distance ladder anchor galaxies, and sociologically may be a palatable resolution for cosmologists,  allowing one a potential resolution to $H_0$ tension if the \textit{local} Universe is not FLRW. To get around this, one could just push the domain of FLRW cosmology to higher redshifts. Alternatively, one could attempt to model the local Universe, but early indications suggest this could make $H_0$ tension worse \cite{Giani:2023aor}.

However, one can caution against this idea. We observe that hemisphere decompositions of both the CMB and Type Ia SNe samples under standard assumptions typically do not lead to pronounced enough angular variations of $\Lambda$CDM parameters to resolve tensions~\cite{Krishnan:2021jmh, Zhai:2022zif, McConville:2023xav, Fosalba:2020gls, Yeung:2022smn}. The key point here is that while angular variations are evident, angular variations in Planck data centre on $H_0 \sim 67 {\rm \,km\, s^{-1}\, Mpc^{-1}}$, whereas angular variations in Cepheid-SNe centre on $H_0 \sim 73 {\rm \,km\, s^{-1}\, Mpc^{-1}}$. That is, the reported values of $H_0$ may be viewed as a values averaged over the sky. Furthermore, even allowing for systematics, or alternatively the potential impact of foregrounds on the CMB \cite{Kovacs:2015hew, DES:2021cge, Lambas:2023gzy} in Planck data, terrestrial CMB experiments clearly agree on a lower $H_0 < 70 {\rm \,km\, s^{-1}\, Mpc^{-1}}$ value~\cite{ACT:2020gnv, SPT-3G:2022hvq}. In summary, while a violation of FLRW in the local Universe $z \lesssim 0.1$ seems compelling~\cite{Aluri:2022hzs}, and can impact the distance ladder~\cite{Mortsell:2021tcx, Perivolaropoulos:2021bds, Perivolaropoulos:2023iqj, Lane:2023ndt}, the prospect of resolving $H_0$ through relaxing FLRW currently seems distant.


Moving along to $S_8$ tension, it is worth stressing that since weak lensing sensitivity peaks at  $z \sim 0.4$ (see Fig.~1 from~\cite{ACT:2023kun}), a local $z \lesssim 0.1$ FLRW violation is not expected to address the $S_8$ tension in weak lensing. Nevertheless, it may explain lower values of $S_8$ from peculiar velocities~\cite{Boruah:2019icj, Said:2020epb, Hollinger:2023mrp} and potentially  Sunyaev-Zeldovich (SZ) cluster counts~\cite{Planck:2013lkt}, since samples are biased to lower redshifts. Moving beyond the local Universe, lower values of $S_8$ observed in redshift space distortions (RSD)~\cite{Macaulay:2013swa,Battye:2014qga, Nesseris:2017vor, Kazantzidis:2018rnb, Skara:2019usd, Quelle:2019vam, Li:2019nux, Benisty:2020kdt, Nunes:2021ipq}, especially evolution of $S_8$ or $\sigma_8$ with redshift~\cite{Esposito:2022plo, Adil:2023jtu} \footnote{There is an independent anomaly in the late-time integrated Sachs-Wolfe (ISW) effect from supervoids \cite{DES:2018nlb, Kovacs:2021mnf}. In particular, an excess ISW effect is reported at lower redshifts \cite{DES:2018nlb}, whereby $A_{\textrm{ISW}} \gtrsim 5$, and a deficit at higher redshifts \cite{Kovacs:2021mnf} with $A_{\textrm{ISW}} \lesssim -5$. $A_{\textrm{ISW}}$ varies with redshift and one encounters the expected $\Lambda$CDM value $A_{\textrm{ISW}} =1$ at intermediate redshifts. Translated into the growth parameter $f$ \cite{Kovacs:2021mnf}, this implies weaker and stronger growth of structures than Planck-$\Lambda$CDM expectations at lower and higher redshifts, respectively. Given the overlap with $f \sigma_8(z)$ constraints from RSD, it is plausible that the ISW anomaly and $S_8$ evolution in RSD \cite{Adil:2023jtu} are symptoms of the Planck-$\Lambda$CDM cosmology being simply an approximation that underestimates or overestimates quantities when one performs a tomographic analysis, but on average gets the right result.}, and evolution in the Weyl potential with DES data~\cite{Tutusaus:2023aux} point to an $S_8$ problem at cosmological scales no less than $z \sim 0.5$, necessitating an unfathomably large scale violation of FLRW. However, if it turns out that the RSD and Weyl potential results are false, and one succeeds in showing no evolution in the $S_8$ parameter down to redshifts of $z \sim 0.2$, as recently claimed~\cite{ACT:2023oei} \footnote{Note this is contradicted by an increasing trend of $S_8/\sigma_8$ in Fig.~10 of \cite{ACT:2024okh} and Fig.~16 of \cite{Sailer:2024coh}, admittedly at low statistical significance.}, one may be able to localise both $H_0$ and $S_8$ tension in the relatively local Universe. That being said, it is clear that one can currently draw different conclusions from different observables, so systematics are clearly an issue.  

The importance of localising missing physics should now be evident to the reader. One could contemplate absorbing both $H_0$ and $S_8$ tensions into a local violation of FLRW. However, if  tensions cannot be localised enough at low redshifts, this idea does not work unless FLRW is violated at large scales {($ z>0.1$)}. As we see it, while radio galaxy and QSO samples used in cosmic dipole studies~\cite{Singal:2011dy, Rubart:2013tx, Singal:2019pqq, Secrest:2020has, Siewert:2020krp, Secrest:2022uvx} must be at high redshift to avoid a spurious clustering dipole~\cite{Tiwari:2015tba}, they are simply null tests of FLRW. Despite allowing FLRW to be violated at redshifts up to $z \sim 1$, they do not narrow down a relevant redshift range. 

\subsection{Matter, Radiation, and $\Lambda$}

We now turn our attention to the additional assumptions defining the $\Lambda$CDM cosmology. A solution to~\eqref{eq:Einstein} subject to the spacetime assumption~\eqref{eq:FLRW}, and the further assumption of vanishing spatial curvature, requires one to specify the stress-energy tensor $T_{\mu \nu}$. Therein lies the physics. We note that one can reconstruct the Hubble parameter ${H:= \dot{a}/a}$ {(where a dot denotes ${\rm d}/{\rm d}t$)} directly from observational Hubble data (OHD)~\cite{Ma:2010mr, Moresco:2015cya, Moresco:2018xdr, Vagnozzi:2020dfn, Jiao:2022aep} or other observational data using Gaussian Process regression~\cite{Holsclaw:2010nb, Holsclaw:2010sk, Shafieloo:2012ht, Seikel:2012uu,Keeley:2020aym}, more general machine learning algorithms~\cite{Nesseris:2012tt, Arjona:2019fwb, Mukherjee:2022yyq, Bengaly:2022cgs, Giambagli:2023ngt,Gomez-Vargas:2022bsm, Dialektopoulos:2023dhb,Medel-Esquivel:2023nov}, {and even wavelets anchored on the comoving distance to the last scattering surface \cite{Akarsu:2022lhx}}, but without a solution to the Einstein field equations, there is no physics. The utility of data reconstructions is that one may diagnose departures from a specific model, e.g., the $\Lambda$CDM model, but care is required with quantifying degrees of freedom (see for example~\cite{Zhao:2017cud, Wang:2018fng,Padilla:2019mgi,Escamilla:2021uoj,Escamilla:2023shf}) and confidence intervals~\cite{OColgain:2021pyh}.\footnote{Interestingly, extrapolations of $H(z)$ reconstructions based on cosmic chronometer data~\cite{Jimenez:2001gg}, a cosmological model agnostic observable, have favoured lower $H_0< 70 {\rm \,km\, s^{-1}\, Mpc^{-1}}$ values~\cite{Busti:2014dua, Gomez-Valent:2018hwc, Haridasu:2018gqm}. However, if systematics are properly propagated, the errors are too large to exclude local $H_0$ determinations~\cite{Moresco:2023zys}, so they do not currently arbitrate on $H_0$ tension.}
  
Returning to specifying $T_{\mu \nu}$, we require matter (humans exist) and radiation (the sun shines). Observationally, matter and radiation are beyond question and we possess a deep theoretical understanding of both. It is a reasonable additional assumption that matter is pressureless at large scales. That being said, one should still test this assumption by checking that any high redshift Hubble parameter scales as $H(z) \sim 100 \sqrt{\Omega_{\rm m}h^2} (1+z)^{\frac{3}{2}}$, {(with $h:= H_0/100\,\rm {\rm \,km\, s^{-1}\, Mpc^{-1}}$ being the reduced Hubble parameter)} in the matter dominated regime. Note, if $\Omega_{\rm m}h^2$ changes with effective redshift, the pressureless matter assumption is rejected. See \cite{Liu:2024fjy} for a recent study in this direction.  

Beyond this point, one runs into stronger assumptions based on weaker supports that are \textit{either} theoretically \textit{or} observationally motivated. Theoretical  preconceptions about a big bang and inflationary epoch have led to the assumption of purely adiabatic scalar primordial perturbations with a nearly scale-invariant spectrum. Observationally, the matter sector is further decomposed into baryonic and dark matter (DM) components, and the discovery of a late-time accelerated expansion~\cite{SupernovaSearchTeam:1998fmf, SupernovaCosmologyProject:1998vns} necessitates an additional DE sector, which is modelled through $\Lambda$. This brings us to the flat $\Lambda$CDM Hubble parameter:  
\begin{eqnarray}
\label{eq:LCDM}
H(z) &=& H_0 \sqrt{ \Omega_{\Lambda} + \Omega_{\rm m} (1+z)^3 + \Omega_{\rm r} (1+z)^4}, \nonumber \\
\Omega_{\Lambda} &=& 1-\Omega_{\rm m} - \Omega_{\rm r}, \quad \Omega_{\rm m} = \Omega_{\rm c} + \Omega_{\rm b}, 
\end{eqnarray}
where $H_0$, $\Omega_{\rm b}$, $\Omega_{\rm c}$, $\Omega_{\rm r}$, and $\Omega_{\rm m}$ are parameters representing the present-day ($z=0$) values of the Hubble expansion rate and the density parameters for baryonic matter, (cold) DM, radiation, and pressureless matter (baryons+DM), respectively. These parameters are mathematically  constants of integration of Friedmann and continuity equations  when these sectors are only gravitationally interacting~\cite{Krishnan:2020vaf}. While integration constants from the mathematical perspective, these  are fitting parameters of the model when confronted with data. Their value read from different data sets at different redshifts may not turn out to be redshift-independent constants. $\Omega_{\rm r}$ is fixed by the CMB monopole temperature, while constraints on $H_0$, $\Omega_{\rm b}$, and $\Omega_{\rm c}$ are derived from Planck CMB data, thereby defining the Planck-$\Lambda$CDM model. 
The mathematical consistency of flat $\Lambda$CDM model then implies/requires constancy of the observationally inferred values of these fitting parameters~\cite{Colgain:2022rxy,Colgain:2022tql}. 

For theoretical physicists, the presence of $\Omega_{\Lambda}$ in (\ref{eq:LCDM}) is deeply perplexing. There are two interpretations in the literature. The most nailed down (testable) interpretation is that $\Lambda$ is the cosmological constant from GR (\ref{eq:Einstein}). If this is indeed the case, one expects to see traces of $\Lambda$ at all scales where gravity is relevant, and in principle, this opens up tests of $\Lambda$ in the solar system~\cite{Kerr:2003bp, Arakida:2011ty, Arakida:2012ya} or local group~\cite{Benisty:2023vbz, Benisty:2023clf, Benisty:2024lsz}. It also begets the longstanding and unsolved cosmological constant problem~\cite{Weinberg:1988cp, Weinberg:2000yb}, which demands a physical explanation to the observed smallness of the cosmological constant in the face of theoretical expectations from quantum physics that it should be much larger. The second interpretation is that $\Omega_{\Lambda}$ is purely a phenomenological term in the flat $\Lambda$CDM model. This then allows one to introduce other phenomenological parametrisations, such as the Chevallier-Polarski-Linder (CPL) model~\cite{Chevallier:2000qy, Linder:2002et}, to test whether DE density is a constant or not. In cosmology, there is a tendency to flit between these two interpretations. Progress here has been slow, and 25 years after the discovery of DE~\cite{SupernovaSearchTeam:1998fmf, SupernovaCosmologyProject:1998vns}, little progress has been made on whether $\Lambda$ is the cosmological constant from GR or not.

Nevertheless, this may be changing with recent large SNe compilations from Pantheon+~\cite{Brout:2022vxf}, Union3~\cite{Rubin:2023ovl} and DES~\cite{DES:2024tys} showing a preference for an evolving DE equation of state. Here, two points are worth stressing. First, CPL is not a model and it is merely a diagnostic for dynamical dark energy, because all one is doing is performing a Taylor expansion in $(1-a)$ in the DE equation of state. As commented upon earlier, one could equally apply a CPL-like ansatz to the matter sector to test if matter is pressureless. Secondly, as remarked in~\cite{Colgain:2021pmf}, CPL is an expansion in a smaller redshift-dependent parameter, $1-a = z/(1+z)$, which makes CPL less sensitive to low redshift data. In other words, care is required to check which redshift ranges are driving deviations from $\Lambda$ in the CPL model~\cite{Rubin:2023ovl, DES:2024tys}. It cannot be naively assumed that data in the traditional DE dominated regime $z \lesssim 0.7$ is responsible for the deviation.


Returning to CMB, the flat $\Lambda$CDM fitting parameters are fixed up to sub-percent errors once one fits Planck data~\cite{Planck:2018vyg}. The effective redshift of the CMB is $z \sim 1100$. Thus, bearing in mind that redshift is a proxy for time, from the perspective of physics one is in effect fixing all the degrees of freedom of the $\Lambda$CDM model using data on a single time-slice. As a result, the $\Lambda$CDM model is superpredictive; the model either leads to consistent results on independent time-slices or it does not. In particular, while late Universe determinations of $\Omega_{\rm m}$ from Type Ia SNe and baryon acoustic oscillations (BAO) are largely consistent with Planck~\cite{Brout:2022vxf, eBOSS:2020yzd}, the problem is that neither $H_0$ nor $S_8$ may be consistent. Neglecting systematics, this is enough to rule out the model, because the Planck-$\Lambda$CDM model is a dynamical model of the Universe with no free degrees of freedom.

\section{Testable $\Lambda$CDM subsectors}
In physics, and science more generally, given a dynamical model with constant fitting parameters and time series data, there is an inevitable question about the range of validity of the model. 
\begin{quote}
\textit{A valid, self-consistent model is a model that returns the same fitting parameters in a given time domain or epoch. If the model does not, and the fitting parameters evolve outside of the errors, then the model is not predictive and is meaningless in a physics context.}
\end{quote}
Note, the time domain where the model is valid may not cover all the time series data, in which case the model is an effective model that is only valid in a restricted regime. Any putative model of the observable Universe is expected to return consistent values of the fitting parameters from recombination at $z \sim 1100$ down to $z \sim 0 $ today. Systematics aside, $\Lambda$CDM tensions caution that this may not be true. Obviously, this is at odds with Bayesian model comparison, where it is inherently assumed that model fitting parameters are constant.\footnote{This becomes a little more puzzling in the context of $H_0$ tension, an apparent serious $> 3 \sigma$ discrepancy. In order to compare model $A$, i.e., $\Lambda$CDM, to model $B$, a potential replacement that resolves/alleviates $H_0$ tension, one has to combine data sets that are inconsistent when confronted to model $A$.}

\subsection{$H_0$ tension subsector}

The standard $\Lambda$CDM model is a 6-parameter model. This leads to complicated degeneracies between the parameters. Nevertheless, one can identify two subsectors of the flat $\Lambda$CDM model where one can safely decouple additional parameters. 

First, the modelling can be greatly simplified by decoupling the radiation sector from (\ref{eq:LCDM}) and removing perturbative physics. The latter is justified in the $H_0$ tension context, as $H_0$ is defined through (\ref{eq:Einstein}) and (\ref{eq:FLRW}). Alternatively put, perturbations impact peculiar velocities, but the rate of expansion of the Universe must not depend on peculiar velocities. The radiation sector is easily decoupled since $\Omega_{\rm r} \ll \Omega_{\rm m}$, and as a result the $\Omega_{\rm r} (1+z)^4$ term in (\ref{eq:LCDM}) can be ignored if one works at lower redshifts, where $z \lesssim 30$ is conservatively low redshift. This reduces the $\Lambda$CDM model at the background level to a 2-parameter model:
\begin{equation}
\label{eq:LCDM_late}
H(z) = H_0 \sqrt{1-\Omega_{\rm m} + \Omega_{\rm m} (1+z)^3}, 
\end{equation}
parametrised exclusively through $H_0$ and matter density $\Omega_{\rm m}$. What is important here is that the Universe's look-back time in gigayears (Gyr) at redshift $z$ is  
\begin{equation}
\label{eq:age}
t(z) = 977.8 \int_0^{z} \frac{\dd z^{\prime} }{(1+z^{\prime}) H(z^{\prime})} \textrm{ Gyr}. 
\end{equation}
Since $H(z)$ is a strictly increasing function of redshift, this means that the $\Lambda$CDM model describes about 13 billion years of evolution at the level of equations of motion with only two parameters $(H_0, \Omega_{\rm m})$. Moreover, $H_0$ is simply an overall scale, as dictated by the FLRW assumption, so it has no bearing on any evolution in time or its proxy redshift. 

The current quality of astronomical data aside, there is good reason to believe the evolution of the Universe cannot be described by (\ref{eq:LCDM_late}) over 13 billion years. Here is the argument. Consider a generic Taylor expansion of the Hubble parameter in $(1-a):=y = z/(1+z)$ about $a=1$,  
\begin{equation}
\label{eq:Taylor}
H(z) = H_0 \sum_{n=0}^{N} b_n y^n,  \qquad y = \frac{z}{1+z},
\end{equation}
where we set $b_0 = 1$ so that $H(z=0) = H_0$. As explained by Catto\"en \& Visser~\cite{Cattoen:2007sk}, in contrast to Taylor expansion in $z$, the validity of the expansion in $y$ is not confined to $z \lesssim 1$. The only assumption being made in~\eqref{eq:Taylor} is that $H(z)$ is a continuous function, but otherwise one is looking at a model with $N$ degrees of freedom. Nevertheless, in the $\Lambda$CDM model~\eqref{eq:LCDM_late}, there is only one degree of freedom $\Omega_{\rm m}$. As a result, if the $\Lambda$CDM model accurately describes the Universe, the $N$ parameters $b_n, 1 \leq n \leq N$, must all depend on a single parameter $\Omega_{\rm m}$. 

\begin{quote} 
\textit{In the limit $N \rightarrow \infty$, and in the absence of a theoretical reason and framework,\footnote{Unlike the firm theoretical guiding principle for particle physics, such as Wilsonian effective field theory, the notion of relevant, marginal and irrelevant operators, decoupling of scales and cluster decomposition, cosmology lacks such guiding principles. These guiding principles make it possible to effectively verify the theoretically important notion of ``stability of description''. That is, only a small number of deformations yield relevant perturbations around a given model, which is a point in the space of all possible models. In the absence of such principles there is no systematical way to classify  deformations relevant to certain features and argue for or verify stability of description in cosmology.} the $\Lambda$CDM model becomes a set of measure zero in the space of all possible Hubble parameters $H(z)$.}
\end{quote}
This is a bet that no discerning scientist should take, but what keeps the $\Lambda$CDM model relevant is the relatively poor quality of cosmological data. Thus, the task remains to show that the $b_n$ do not converge to their $\Lambda$CDM expectations. While this is not a feasible exercise with a large number of parameters $b_n$,  one can expedite the process by simply fitting the model~\eqref{eq:LCDM_late} to observational data in redshift bins and looking for (likely) evolution of the fitting parameter $\Omega_{\rm m}$ with redshift. {Note, if $\Lambda$CDM is falsified, one cannot guarantee that minimal extensions of the $\Lambda$CDM model will turn out to be valid models.\footnote{{For example, consider the $w$CDM model as a 1-parameter extension of the $\Lambda$CDM model, with a constant DE equation of state $w$ that can be viewed as a deformation of $\Lambda$CDM. In keeping with the general thrust of the paper, there is of course no guarantee that $w$ is a constant at different redshifts, e.g.~\cite{Dong:2023jtk}. However, \textit{even in the same redshift ranges}, one may find contradictory results. For example, Union3 and DES SNe have a preference for $w > -1$~\cite{Rubin:2023ovl, DES:2024tys}, whereas SPT Clusters with DES/HST weak lensing prefers $w < -1$~\cite{DES:2024zpp}. Systematics aside, this says that $w$ is not a good deformation parameter around $\Lambda$CDM. Obviously, if the fitting parameters evolve, none of these models are good physical models.}}} 

\subsection{$S_8$ tension subsector}
While the $\Lambda$CDM subsector (\ref{eq:LCDM_late}) is apt for studying $H_0$ tension, one can define a comparable subsector for $S_8$ tension.  To that end, we re-emphasise that redshift is more fundamental than scale. This is a simple observation, but it is routinely overlooked. In short, one can only define density perturbations $\delta(r, t)$ once one specifies a Hubble parameter $H(z)$, i.e., a solution to the Einstein field equations (\ref{eq:Einstein}), and adopts redshift as a proxy for time, $a:= (1+z)^{-1}$. Scale $k$ only enters once one defines a perturbation $\delta$ in matter density, $\rho_{\rm m} = \bar{\rho}_{\rm m} (1 + \delta)$, where $\bar{\rho}_{\rm m}$ denotes the average matter density and we assume the perturbation is small, $\delta \ll 1$. A standard analysis in perturbation theory that combines the continuity, Euler and Poisson equations (Newtonian gravity) then leads to
\begin{equation}
\label{eq:pert1}
\left( \frac{\partial^2}{\partial t^2} + 2 H \frac{\partial}{\partial t} - \frac{c_{\rm s}^2}{a^2} \nabla^2 - 4 \pi G \bar{\rho}_{\rm m} \right) \delta = 0, 
\end{equation}
where $c_{\rm s}$ is the speed of sound. The scale $k$ then enters when one decomposes $\delta$ in Fourier modes, $\nabla^2 \delta = k^2 \delta$. However, provided one works on large scales, $k \ll \sqrt{4 \pi G \bar{\rho}_{\rm m}} (a/c_{\rm s})$, one can safely neglect $k$. In this limit,~\eqref{eq:pert1} reduces to 
\begin{equation}
\label{eq:pert2}
\ddot{\delta} + 2 H \dot{\delta} - 4 \pi G \bar{\rho}_{\rm m} \delta = 0,  
\end{equation}
where dots denote derivatives with respect to time, and $H$ and $\bar{\rho}_{\rm m} := 3 H_0^2 \Omega_{\rm m}/(8 \pi G) a^{-3}$ are defined through the scale factor $a$ at the background level. The key point is that provided one works at linear scales, typically $0.001 h \textrm{ Mpc}^{-1} \lesssim k \lesssim 0.1 h \textrm{ Mpc}^{-1}$, then scale $k$ is irrelevant. {Moreover, at these scales DE is smooth, so that in the $\Lambda$CDM model the same parameter $\Omega_{\rm m}$ determines both background and perturbed evolution. This reinforces the importance of $\Omega_{\rm m}$ across billions years of cosmological evolution.} Unfortunately, the reality is that most cosmological observables probe physics beyond these linear scales (see~\cite{Huterer:2022dds} for a deeper discussion), so great care is required to make sure that one is working in a strict linear regime.  

\section{Localising $\Lambda$CDM tensions}
The basic approach to $\Lambda$CDM tensions, in particular $H_0$ and $S_8$ tension, advocated in this work is to reduce the $\Lambda$CDM model to subsectors with a minimal number of fitting parameters. Once this is done, one can bin data by effective redshift and confront it to simplified settings~\eqref{eq:LCDM_late} and~\eqref{eq:pert2}. 

\begin{quote}
\textit{We stress that since redshift is defined at the background level, but scale is defined at the level of perturbations, one needs to study redshift evolution of fitting parameters before their scale evolution. Moreover, redshift evolution implies a scale problem, but the converse is not true.}
\end{quote} 
Naturally, we understand why the cosmology community may like to focus on scale, as it is less disruptive to the $\Lambda$CDM model, but Nature can be fickle. The goal then is to identify redshift ranges where the $\Lambda$CDM model returns consistent fitting parameters and redshift ranges where the parameters are no longer consistent and evolve or drift with redshift. The latter redshifts are then candidates for missing physics, i.e., physics not  incorporated by the $\Lambda$CDM model. We review the current status of these consistency checks of the model. We begin with $S_8$ tension because the progress is more tangible, and from our perspective $S_8$ tension is more appealing, since it has some semblance of problem that can be addressed with novel scale dependent physics, e.g., baryonic feedback~\cite{Amon:2022azi, Preston:2023uup}. This encourages cosmologists to engage with the $S_8$ problem, while it is not yet as statistically significant as $H_0$ tension. Nevertheless, as stressed, $H_0$ tension is not an obvious scale problem.   

\subsection{Localising $S_8$ tension}
Over the last year, noticeable progress has been made on $S_8$ tension. While $S_8$ tension is primarily seen as a $2-3 \sigma$ discrepancy between lower $S_8$ values from weak lensing~\cite{Heymans:2013fya, Joudaki:2016mvz, DES:2017qwj, HSC:2018mrq, KiDS:2020suj, DES:2021wwk} and a larger Planck value~\cite{Planck:2018vyg}, ACT CMB lensing recently localised the problem (if there is a problem) to the late Universe~\cite{ACT:2023kun, ACT:2023dou, ACT:2023ipp}. The key point is that despite CMB lensing being sensitive to larger redshifts (see Fig. 1 of~\cite{ACT:2023kun}) and larger scales than weak lensing, ACT independently recovers the Planck $S_8$ value. 

\begin{quote}
\textit{One can now define $S_8$ tension as a putative problem in the redshift range $z \lesssim 2$.}
\end{quote} 
This interpretation is supported by independent observations, including lower $S_8$ inferences from peculiar velocities in the local Universe~\cite{Boruah:2019icj, Said:2020epb, Hollinger:2023mrp}, SZ cluster counts~\cite{Planck:2013lkt} and RSD~\cite{Macaulay:2013swa,Battye:2014qga, Nesseris:2017vor, Kazantzidis:2018rnb, Skara:2019usd, Quelle:2019vam, Li:2019nux, Benisty:2020kdt, Nunes:2021ipq}. Moreover, separately it has been noted that $S_8$, or more accurately $\sigma_8$, may evolve from lower to higher values between low and high redshifts in these observables~\cite{Esposito:2022plo, Adil:2023jtu, Tutusaus:2023aux}. If true, this provides independent corroborating evidence for the discrepancy between weak and CMB lensing. In contrast, a recent study~\cite{ACT:2023oei} of cross-correlations of unWISE galaxies and CMB lensing is in conflict with these observations and recovers the Planck value at redshifts $z\gtrsim 0.2$. More recently, a study of cross-correlations of DESI luminous red galaxies with CMB lensing \cite{ACT:2024okh, Sailer:2024coh} has shown that i) $S_8$ or $\sigma_8$ exhibits a mild increasing trend with bin redshift, ii) $S_8$ or $\sigma_8$ is robust to scale cuts, and iii) baryonic feedback cannot explain the results, since the study is primarily sensitive to the linear regime. Note, the $S_8$ literature is extremely confusing in the sense that there is no clear pattern emerging from results. $S_8$ whisker plots are everywhere, but there is a focus on scale-dependent physics, e. g. baryonic feedback \cite{Amon:2022azi, Preston:2023uup}, whereas our arguments here point to missing redshift-dependent physics, in essence because one cannot resolve $H_0$ tension with scale-dependent physics. Interestingly, at redshifts beyond $z \sim 2$ one finds reports of lower values of $\sigma_8$ relative to Planck, which are ostensibly at odds with ACT results~\cite{Miyatake:2021qjr} (see also~\cite{Alonso:2023guh}). Evidently, systematics are not fully under control. 

The relevant question now is whether evolution of $S_8$ or $\sigma_8$ with scale $k$ could be mimicking evolution of $S_8$ or $\sigma_8$ with redshift. While it is true that weak lensing probes smaller scales than CMB lensing, peculiar velocities lead to lower $S_8$ values while still probing larger scales~\cite{Boruah:2019icj, Said:2020epb, Hollinger:2023mrp}. Moreover, both galaxy cluster number counts and Lyman-$\alpha$ spectra, analysed in~\cite{Esposito:2022plo}, are expected to be probing similar scales. In addition, evolution of $\sigma_8$ with redshift is evident across $f \sigma_8(z)$ constraints from RSD and all data are expected to be probing linear scales~\cite{Adil:2023jtu}.\footnote{In recent years, RSD has been extended to non-linear scales in order to extract better constraints from the data. This leads to lower values of $f \sigma_8(z)$, as is evident from Fig. 7~\cite{Chapman:2023yuq}, so scale certainly impacts RSD constraints.} Finally, in~\cite{Tutusaus:2023aux} the Weyl potential is probed through an observable that is expected to be independent of scale, nevertheless once specialised to the $\Lambda$CDM setting, differences in $\sigma_8$ are reported across weak lensing bins in DES data~\cite{DES:2021bpo}.\footnote{As demonstrated~\cite{Tutusaus:2023aux}, scale cuts inflate errors, but have little bearing on the central values. One can of course trivially resolve $S_8$ tension by throwing information away.} Given these results, while it may be appealing to the cosmology community if $S_8$ increased with increasing scale, this is not borne out in the data. That being said, it is reasonable to expect some $S_8$ constraints in the literature are scale dependent.     

At no more than $2-3 \sigma$ statistical significance, and potentially even less significant~\cite{Kilo-DegreeSurvey:2023gfr}, the jury is still out on whether $S_8$ tension is a definite problem, however the fact that we see a lower $S_8$ or $\sigma_8$ in at least 4 independent observables, i.e., weak lensing, peculiar velocities, RSD, cluster counts, which probe lower redshifts points to an obvious problem. The fact that higher redshift probes recover the Planck $S_8$ value offers support to the idea that $S_8$ tension is due to missing physics in the late Universe, which conservatively can be demarcated as redshifts $z \lesssim 2$. This begets an interesting question. Assuming there is missing physics at the perturbative level, and given that $H_0$ tension is defined at the background level, how is one sure that there is no missing physics at the background level? Note, while evolution of $S_8$ with scale can be addressed with new perturbative physics, evolution of $S_8$ with redshift is a different kettle of fish, forcing one also to change the background. This nicely brings us to $H_0$ tension and consistency tests investigating the constancy of the $\Lambda$CDM fitting parameter $H_0$ from (\ref{eq:LCDM_late}).  

\subsection{Localising $H_0$ tension}
Here we focus on how progress can be made on $H_0$ tension. First, we need to confirm that local (cosmological model independent) $H_0$ determinations consistently return larger $H_0 > 70 {\rm \,km\, s^{-1}\, Mpc^{-1}}$ values. Combined with the robustness of early Universe determinations across CMB experiments~\cite{Planck:2018vyg, ACT:2020gnv, SPT-3G:2022hvq} and BBN+BAO~\cite{Addison:2017fdm, Cuceu:2019for, Schoneberg:2019wmt, Schoneberg:2022ggi}, this demonstrates that there must be missing physics in the $\Lambda$CDM model, not just at a perturbative level, as outlined above, but also at the background level. 

Recent observations suggest that we are well on our way to achieving a $H_0 > 70 {\rm \,km\, s^{-1}\, Mpc^{-1}}$ local measurements. In particular, Cepheid-SNe $H_0$ inferences have exhibited stable central values of $H_0 \sim 72-74 {\rm \,km\, s^{-1}\, Mpc^{-1}}$ over the last decade~\cite{HST:2000azd, Freedman:2012ny, Follin:2017ljs, Cardona:2016ems, Riess:2021jrx}. Importantly, independent groups agree on the result and JWST has thrown up no major surprises on Cepheid calibration~\cite{Riess:2023bfx}. Moreover, a host of independent distance indicators, including megamasers~\cite{Pesce:2020xfe}, surface brightness fluctuations~\cite{Blakeslee:2021rqi}, Tully-Fisher relation~\cite{Kourkchi:2020iyz, Schombert:2020pxm} and Type II SN~\cite{deJaeger:2020zpb, deJaeger:2022lit} have independently led to $H_0 > 70 {\rm \,km\, s^{-1}\, Mpc^{-1}}$.\footnote{We note that a gravitational wave standard siren $H_0$ determination exists~\cite{LIGOScientific:2017adf, Nicolaou:2019cip}, but with only one event, the errors are not currently competitive.} Tip of the Red Giant Branch (TRGB) calibration has regularly returned lower values in this window, leading to considerable debate~\cite{Reid:2019tiq, Freedman:2020dne, Freedman:2021ahq, Anderson:2023aga, Scolnic:2023mrv}, but there appears to be consensus now that $H_0 > 70 {\rm \,km\, s^{-1}\, Mpc^{-1}}$~\cite{Uddin:2023iob}. The only bone of contention is whether $1 \%$ errors are achievable.\footnote{Ref.~\cite{Uddin:2023iob} originally pointed to larger systematic errors, but the errors have since been revised downwards. One eventually expects to reach a precision where hemisphere decompositions, or equivalent, of distance indicators lead to differences in $H_0$ values on the sky that exceed the errors~\cite{Zhai:2022zif, McConville:2023xav}, which is expected if the Universe at lower redshifts is not FLRW. }

In some sense, we appear to have already confirmed locally that $H_0$ is in $70-77 {\rm \,km\, s^{-1}\, Mpc^{-1}}$ range, and all that remains is to debate the errors. The next exercise of interest is to study $H_0$ determinations at cosmological scales where \textit{one must assume a model}. Invariably, the assumption made is that the $\Lambda$CDM model is correct, which alarmingly is only the case if there is no $H_0$ tension!  
\begin{quote}
\textit{If $H_0$ tension is physical, and the $\Lambda$CDM model has broken down, it is reasonable to expect $H_0$ inferences at cosmological scale to be biased if they assume $\Lambda$CDM. Not allowing for this possibility runs the risk of circular logic. } 
\end{quote}

For this reason, one may expect $H_0$ determinations from strong lensing time delay~\cite{Wong:2019kwg, DES:2019fny, Millon:2019slk,  Yang:2020eoh, Birrer:2020tax, Denzel:2020zuq, Shajib:2023uig} and gravitational wave dark sirens~\cite{DES:2020nay, Palmese:2021mjm, DESI:2023fij} to be potentially impacted by model breakdown, if the model is breaking down in the late Universe.  Whether the former converges to a $H_0$ value consistent with local or early Universe $H_0$ determinations depends on the assumptions being made~\cite{Birrer:2020tax}. Recently, a strongly lensed Type Ia supernova has led to a $H_0$ value~\cite{Kelly:2023mgv} that favours a $H_0$ value consistent with early Universe determinations, so evidently lensed systems have not been able to arbitrate $H_0$ tension.\footnote{A real concern here is that if there is a large spread in $H_0$ values from strong lensing with constrained errors assuming the $\Lambda$CDM model, one arrives at the conclusion that strong lensing cannot consistently determine $H_0$. Ultimately, it is plausible that lens degeneracies, most notably the mass sheet transformation \cite{MST}, make it difficult to determine $H_0$ uniquely \cite{Wagner:2018rae, Wagner:2019azs}. Nevertheless, if one succeeds in precluding systematics, such an outcome could rule out the $\Lambda$CDM model \cite{Li:2024elb}.} On the other hand, dark siren constraints are not yet strong enough to distinguish between local and early Universe $H_0$ determinations.    

One situation where we may have seen a hint of model breakdown is H0LiCOW and TDCOSMO reported a descending trend of $H_0$ with lens redshift, admittedly at low statistical significance $\sim 1.7 \sigma$~\cite{Wong:2019kwg, DES:2019fny, Millon:2019slk}. What matters here is that this low significance trend is not an obvious systematic~\cite{Millon:2019slk}. Moreover, the lensed system driving this trend at the lowest redshift probed was recently revisited by analysing spatially resolved stellar kinematics~\cite{Shajib:2023uig}, and while the errors have inflated, the central value $H_0 \sim 77 {\rm \,km\, s^{-1}\, Mpc^{-1}}$ is unchanged from~\cite{Wong:2019kwg}. Given that TDCOSMO plan to release 40 odd $H_0$ determinations from different lensed QSOs with different lens redshifts in future, if there is a trend, this will hopefully be evident. 

While each strongly lensed QSO or supernova system constitutes its own redshift bin, one can look for corroborating evidence by binning independent observables by effective redshift. Concretely, one assumes the model (\ref{eq:LCDM_late}) is correct, bins observables by redshift and studies the fitting parameter $H_0$ with varying effective redshift. This consistency test of a constant $H_0$ fitting parameter was initiated in~\cite{Krishnan:2020obg} with a combination of local and cosmological data, including megamasers, SNe, cosmic chronometer and BAO data. An independent descending trend with statistical signifiance $\sim 2 \sigma$ was reported. It was subsequently noted in~\cite{Dainotti:2021pqg, Dainotti:2022bzg}  that this trend in a combination of data could be recovered through an ansatz when one isolated SNe and BAO data on their own. The result has since been confirmed by a number of different groups with different methodologies and perspectives~\cite{Colgain:2022nlb, Colgain:2022rxy, Hu:2022kes, Jia:2022ycc, Dainotti:2023yrk} (see \cite{Wagner:2022etu} for further discussion). 

The result is curious for at least three reasons. First, we are now looking at distinct decreasing $H_0$ trends with effective redshift assuming the correctness of the $\Lambda$CDM model across different observables that share no obvious systematic, e.g., strongly lensed QSOs, SNe and observational Hubble data (BAO). Secondly, bearing in mind that local Universe $H_0$ measurements~\cite{Riess:2021jrx, Freedman:2021ahq, Pesce:2020xfe, Kourkchi:2020iyz, Blakeslee:2021rqi} are biased to larger values than early Universe cosmological $H_0$ determinations~\cite{Planck:2018vyg}, a decreasing $H_0$ trend with effective redshift is consistent with what one sees in the $H_0$ tension problem. Thirdly, just as the case in strongly lensed QSOs~\cite{Millon:2019slk}, one may struggle to find a systematic in Type Ia SNe that could cause $H_0$ to evolve with effective redshift, especially given that SNe as distance indicators have been studied since the early 1990s~\cite{Phillips:1993ng}. 

Nevertheless, a decreasing $H_0$ with increasing effective redshift trend may only be half the story. The reason being that $\Omega_{\rm m}$ is anti-correlated with $H_0$ in the model (\ref{eq:LCDM_late}). Thus, one expects any decreasing $H_0$ trend with effective redshift to be complemented by an increasing trend of $\Omega_{\rm m}$ with effective redshift. Note, it is possible that the combination $\Omega_{\rm m} h^2$ remains a constant, but this is not {observationally required; as explained, changes in $\Omega_{\rm m}h^2$ imply a problem with the pressureless matter assumption (see \cite{Liu:2024fjy}).} Indeed, building on observations of larger than expected matter density values in standardisable QSOs~\cite{Risaliti:2015zla, Risaliti:2018reu, Lusso:2020pdb} (see also~\cite{Yang:2019vgk}), it was shown in~\cite{Colgain:2022nlb} that standardisable QSOs and Type Ia SNe share two key traits. First, both probes recover a Planck value for $\Omega_{\rm m}$ at lower redshifts $z \lesssim 0.7$, but an increasing trend of $\Omega_{\rm m}$ with redshift is seen at higher redshifts. Upgrading the Pantheon SNe sample~\cite{Pan-STARRS1:2017jku} to the Pantheon+ SNe sample~\cite{Brout:2022vxf} does not eliminate the evolution~\cite{Malekjani:2023ple} \footnote{The same increasing $\Omega_m$ trend with effective redshift can be demonstrated in DES SNe \cite{Colgain:2024ksa}, where there is no overlap with Pantheon+ SNe at higher redshifts \cite{DES:2024tys}.}. Seemingly consistent results with SNe and/or QSOs were reported elsewhere~\cite{Pourojaghi:2022zrh, Pasten:2023rpc}. It is worth stressing that despite this agreement, the standardisability of QSOs is open to debate~\cite{Khadka:2020vlh, Khadka:2020tlm, Dainotti:2022rfz, Singal:2022nto, Petrosian:2022tlp, Zajacek:2023qjm}. Throughout these studies it is observed that changes in $\Omega_{\rm m}$ are noticeable at higher redshifts, either at or beyond the deceleration-acceleration transition redshift $z \sim 0.7$. This highlights a potential problem with the $\Lambda$CDM model, but this problem is only evident at redshifts where the pressureless matter assumption is important. This implies missing physics in the matter sector of the $\Lambda$CDM model.

Nevertheless, there are other angles worth mentioning. JWST has reported high redshift galaxies~\cite{Adams:2022, Labbe:2022, Castellano:2022, Naidu:2022, Xiao:2023} that appear anomalous if the Planck-$\Lambda$CDM model is correct. Nevertheless, it is easy to show that increasing $\Omega_{\rm m} h^2$ in the $\Lambda$CDM model would make preliminary JWST results less anomalous~\cite{Boylan-Kolchin:2022kae}. From Table 1 of~\cite{Colgain:2022nlb}, one can convince oneself that $\Omega_{\rm m} h^2$ is increasing with effective redshift in the QSO sample. This coincidence is intriguing and warrants further study. {Moreover, both Union3 and DES now have a preference for larger than expected $\Omega_{\rm m}$ determinations assuming the $\Lambda$CDM model~\cite{Rubin:2023ovl, DES:2024tys}, thereby throwing the constancy of $\Omega_{\rm m}$ into question. At $\Omega_{\rm m} = 0.356^{+0.028}_{-0.026}$~\cite{Rubin:2023ovl} and $\Omega_{\rm m} = 0.352 \pm 0.017$~\cite{DES:2024tys} respectively, one is looking at $1.5 \sigma$ and $2 \sigma$ differences with Planck, $\Omega_{\rm m} = 0.315 \pm 0.007$~\cite{Planck:2018vyg}. It will be interesting to see if $\Omega_{\rm m}$ is constant across these large SNe samples when the data is binned by redshift.}  Finally, since we see a decreasing $H_0$/increasing $\Omega_{\rm m}$ trend with effective redshift in multiple independent observables, it is relatively easy to combine observables to find a statistically significant result $\gtrsim 3 \sigma$~\cite{Colgain:2022rxy}. Physically, one may try to interpret the results as motivation for a Mpc-scale \cite{Shanks:2018rka, Kenworthy:2019qwq, Haslbauer:2020xaa, Cai:2020tpy, Camarena:2022iae, Mazurenko:2023sex} or Gpc-scale void \cite{Ding:2019mmw, Haslbauer:2023vyf}, but if a model is breaking down, this can be due to a lack of model complexity and there may be no deep physical reason.

It should be stressed that a decrease in $H_0$ compensated by an increase in $\Omega_{\rm m}$ is often misinterpreted as an artefact of the degeneracy or anti-correlation between the two parameters. Indeed, for exclusively high redshift observational Hubble data constraints $H(z)$ or angular diameter/luminosity distance constraints $D_{A}(z)/D_{L}(z)$, one expects 2D Markov Chain Monte Carlo (MCMC) posteriors corresponding to banana-shaped contours in the $(H_0, \Omega_{\rm m})$-plane. Note, the $\Lambda$CDM parameters are confined to $H_0^2\Omega_{\rm m}= \textrm{const.}$ curves in the $(H_0, \Omega_{\rm m})$-plane when one fits exclusively high redshift data~\cite{Colgain:2022tql}, so the observation of banana-shaped 2D MCMC posteriors is expected.\footnote{Observational Hubble data  constrain best fits to the curve $\Omega_{\rm m} h^2 = c$, whereas angular diameter/luminosity distance constraints confine best fit parameters to the curve $(1-\Omega_{\rm m}) h^2 = c'$, where $c,c'$  are data-dependent constants. $c,c'$ may also evolve with effective redshift and its constancy assuming the $\Lambda$CDM model needs to be observationally checked.} These contours are interpreted as the fit to the data from any point in $(H_0, \Omega_{\rm m})$ parameter space is more or less the same. Translated into a frequentist framework, if this is true, one expects to see a curve of almost constant $\chi^2$ that includes the global minimum of the $\chi^2$. Nevertheless, by studying profile distributions~\cite{Gomez-Valent:2022hkb, Colgain:2023bge}, a variant of profile likelihoods~\cite{Herold:2021ksg, Holm:2023laa, Holm:2023uwa}, it is possible to show that points in parameter space that are equivalent from the perspective of MCMC posteriors are no longer equivalent when one analyses the $\chi^2$. In other words, the data distinguishes between points in model parameter space, but this distinction is not evident in MCMC posteriors. We observe that if these trends in the fitting parameters are established, there is evidently physics missing from the $\Lambda$CDM model in the late Universe.  

\section{Outlook}
Given the $H_0$ and $S_8$ discrepancies in the $\Lambda$CDM model~\cite{Perivolaropoulos:2021jda, Abdalla:2022yfr}, it is scientifically prudent to conduct consistency checks of the $\Lambda$CDM model to either confirm model breakdown or assure that unexplored systematics are at play. Given that redshift is more fundamental than scale, since redshift is a proxy for time and the Universe is dynamical, the most relevant consistency checks probe redshift. Simply put, one bins observations by effective redshift, fits the $\Lambda$CDM model and identifies the flat $\Lambda$CDM fitting parameters preferred by the model in each bin. If these fitting parameters are the same at all redshifts within $2 \sigma$, as judged by the errors, then the model passes the consistency check. On the contrary, if the fitting parameters evolve or drift outside of $2 \sigma$, and one sees the same feature in multiple observables,\footnote{This is a necessity in cosmology as it would be foolhardy to trust any single observable. To put this in the context, one only trusts the existence of DE because we see it in multiple observables.} then the model has broken down. Note, the arguments being run are largely targeting modeling 101, namely fitting parameters should not evolve. The $\Lambda$CDM model may also be targeted using more physical arguments, e. g. \cite{Kroupa:2023ubo}, but here our arguments are more mathematical than physical.

Note, in contrast to Bayesian model comparison, there is no second model to compare. Moreover, there are no additional parameters. If the fitting parameters evolve, this worsens the fit to data, and as data improves, one will eventually be able to construct a new model that incorporates the evolution. This is how one eventually connects to Bayesian methods. The key point here is that cosmology purports to be a sub-branch of physics, however physics demands that models are predictive, in our context that is,  fitting parameters should not evolve. We remark that this discussion applies to all cosmological models. They are not specific to the $\Lambda$CDM model. For any model of the Universe to be physically meaningful, it is imperative that the fitting parameters do not change with redshift or scale.

In this work we reviewed the assumptions underlying the $\Lambda$CDM model. We concluded that the foundations based on GR and FLRW do not appear to offer a compelling direction to resolve the tensions. We dispel the first possibility on the grounds that there are simply no observed deviations from GR in controlled environments, where it should be stressed that little in cosmology is under control, thus the need for multiple probes. In contrast, FLRW looks shaky~\cite{Aluri:2022hzs}, especially at lower redshifts $z \lesssim 0.1$, but the angular variations seen in the distance ladder and CMB (they exist!) are simply too small to explain a $\sim 10 \%$ $H_0$ discrepancy~\cite{Krishnan:2021jmh, Zhai:2022zif, McConville:2023xav, Fosalba:2020gls, Yeung:2022smn}. In short, we believe that the FLRW assumption is worth studying, but we do not see it coming to the rescue on $\Lambda$CDM tensions, other than relaxing FLRW leads to $\Lambda$CDM fitting parameters, e.g., $H_0, \Omega_{\rm m}$ and $S_8$, being ill-defined.  

In the latter part of this work we outlined simplified subsectors of the flat $\Lambda$CDM model where one can decouple additional parameters. The $H_0$ tension subsector (\ref{eq:LCDM_late}) allows one to probe the constancy of the fitting parameters $H_0$ and $\Omega_{\rm m}$ across 13 billion years of evolution of the Hubble parameter $H(z)$. Since $H_0$ is a scale, this is 13 billion years modelled by a single parameter $\Omega_{\rm m}$. As argued earlier, (\ref{eq:LCDM_late}) is a set of measure zero in all the continuous functions $H(z)$ that one can construct, thereby making it likely that a deviation will be found as data quality improves. {Union3 and DES may already be hinting at this result~\cite{Rubin:2023ovl, DES:2024tys}.} On the other hand, performing consistency checks of $S_8$ involves restricting observables to linear scales $0.001 h \,\textrm{Mpc}^{-1} \lesssim k \lesssim 0.1 h \, \textrm{Mpc}^{-1}$ where perturbations in matter density are independent of $k$ (\ref{eq:pert2}). Nevertheless, observationally, one needs to make sure that non-linear physics is not impacting observables and this is admittedly tricky.  

Over the last year, progress appears to have been made localising $S_8$ tension to the late Universe, but this may have gone under the radar. Comparison between weak~\cite{Heymans:2013fya, Joudaki:2016mvz, DES:2017qwj, HSC:2018mrq, KiDS:2020suj, DES:2021wwk} and CMB lensing~\cite{ACT:2023kun, ACT:2023dou, ACT:2023ipp} now restricts deviations of $S_8$ from the Planck-$\Lambda$CDM value to $z \lesssim 2$. This raises the possibility that $S_8$ either increases with redshift or scale. Separately, peculiar velocities~\cite{Boruah:2019icj, Said:2020epb, Hollinger:2023mrp}, galaxy cluster counts~\cite{Planck:2013lkt} and RSD favour lower values of $S_8$~\cite{Macaulay:2013swa,Battye:2014qga, Nesseris:2017vor, Kazantzidis:2018rnb, Skara:2019usd, Quelle:2019vam, Li:2019nux, Benisty:2020kdt, Nunes:2021ipq} that are consistent with weak lensing. A scale dependent $S_8$~\cite{Amon:2022azi, Preston:2023uup} can plausibly explain discrepancies between weak and CMB lensing, but it would struggle to explain lower $S_8$ values from peculiar velocities as the determinations are made at large scales. Moreover, three independent studies~\cite{Esposito:2022plo, Adil:2023jtu, Tutusaus:2023aux} incorporating galaxy cluster counts, RSD and weak lensing data now point to variations in $\sigma_8$ with redshift. While this is countered elsewhere~\cite{ACT:2023oei} (however see \cite{ACT:2024okh, Sailer:2024coh}), it behoves us to once again remind the reader that redshift is more fundamental than scale. In addition, the more serious $H_0$ tension problem can only be due to redshift evolution if the $\Lambda$CDM model is breaking down~\cite{Krishnan:2020vaf}. The simplicity of $2+2=4$ cannot be denied. For this reason, a redshift dependent $S_8$ or $\sigma_8$ is a natural and plausible signature of $\Lambda$CDM model breakdown.

This brings us nicely to $H_0$ tension~\cite{Riess:2021jrx, Freedman:2021ahq, Pesce:2020xfe, Kourkchi:2020iyz, Blakeslee:2021rqi}. As explained in~\cite{Krishnan:2020vaf}, a prerequisite to the tension not being due to systematics is a $H_0$ fitting parameter that evolves with redshift in the $\Lambda$CDM model. Mathematics demands that this must be observed, although the redshift range is uncertain and needs localisation. To that end, $H_0$ inferences from strong lensing time delay, Type Ia SNe, standardisable QSOs and OHD all point to a decreasing trend of $H_0$ with increasing effective redshift~\cite{Wong:2019kwg, Millon:2019slk, Krishnan:2020obg, Dainotti:2021pqg, Dainotti:2022bzg, Colgain:2022nlb, Colgain:2022rxy,  Hu:2022kes, Jia:2022ycc, Dainotti:2023yrk}. The statistical significance in each observable may be low $\lesssim 2 \sigma$, but combining results increases the significance quickly; it has been estimated as no less than $3 \sigma$~\cite{Colgain:2022rxy}. On the flip side, as argued earlier, one should not expect to be able to model 13 billion years of background evolution with a single parameter $\Omega_{\rm m}$ dialled to $\Omega_{\rm m} \sim 0.3$. Nevertheless, since $H_0$ and $\Omega_{\rm m}$ are anti-correlated, and one typically encounters banana-shaped contours, i.e., $H_0^2\Omega_{\rm m} =  \textrm{const.}$ curves in the $(H_0,\Omega_{\rm m})$-plane in high redshift bins, because the $\Lambda$CDM model only well constrains combinations of $H_0$ and $\Omega_{\rm m}$ at high redshifts, great care is required in estimating errors and assigning statistical significance. This is a technical problem that needs to be overcome. What is safe to say currently is that banana-shaped 2D MCMC posteriors from the Bayesian approach \textit{do not imply} curves of constant $\chi^2$  in the frequentist approach~\cite{Colgain:2023bge}. Thus, what appear to be equivalent points in $(H_0, \Omega_{\rm m})$-parameter space for Bayesians may no longer be equivalent points in parameter space for frequentists. These conceptual differences will need to be overcome, otherwise frequentists will make discoveries well before the results are adopted by Bayesians. 

At a conceptual level what is being said is simple. If in response to $\Lambda$CDM tensions one builds a replacement model by injecting new physics at a given cosmic epoch, this is only well motivated if it correlates with missing physics in the $\Lambda$CDM model. Thus, whether one confronts the new model to data or the original $\Lambda$CDM model to the same data, one should agree on the redshift ranges or scales where new physics is required. The problem here is that EDE \cite{Poulin:2018cxd, Agrawal:2019lmo, Lin:2019qug, Niedermann:2019olb, Ye:2020btb, Poulin:2023lkg} injects new physics in epochs where there is no data allowing direct tests of the $\Lambda$CDM model. Nevertheless, if evolution of $\Lambda$CDM parameters $H_0$ and $S_8$ is established in the late Universe, one can attempt to correlate this with coupled DE-matter models \cite{Wang:2016lxa, DiValentino:2017iww, DiValentino:2019ffd, Gomez-Valent:2020mqn, Wang:2024vmw}, sign-switching cosmological constant ($\Lambda_{\rm s}$CDM) models \cite{Akarsu:2019hmw,Akarsu:2021fol, Akarsu:2022typ, Akarsu:2023mfb}, etc. 

In summary, $H_0$ tension is a serious anomaly. Given that we only see anomalous results in the local Universe, interpreting it cosmologically is challenging. Nevertheless, the less significant $S_8$ tension promises to be insightful, since we appear to have localised the problem to the late Universe $z \lesssim 2$. From there, it all depends on whether $S_8$ or $\sigma_8$ varies with redshift or scale, but one must preclude redshift first as it is more fundamental. A number of results now favour redshift evolution of $S_8$ or $\sigma_8$, when assuming the $\Lambda$CDM model~\cite{Esposito:2022plo, Adil:2023jtu, Tutusaus:2023aux}. If so, this is a problem that can only be addressed by changing cosmology at the background level. The same is true for the $H_0$ tension problem. Going forward, establishing that the $H_0$ and $\Omega_{\rm m}$ $\Lambda$CDM fitting parameters evolve with redshift in the late Universe should be enough to settle the issue provided we see similar evolution across independent cosmological probes. Preliminary results in that direction exist.

\begin{acknowledgments}
We would like to thank Yashar Akrami, Luca Amendola, Adam Riess, Istv\'an Szapudi and Sunny Vagnozzi for fruitful discussions. \"{O}A acknowledges the support of the Turkish Academy of Sciences in the scheme of the Outstanding Young Scientist Award (T\"{U}BA-GEB\.{I}P). The work of MMShJ is supported in part by INSF grant No 4026712. AAS acknowledges the funding from SERB, Govt of India under the research grant no: CRG/2020/004347. This article/publication is based upon work from COST Action CA21136 – “Addressing observational tensions in cosmology with systematics and fundamental physics (CosmoVerse)”, supported by COST (European Cooperation in Science and Technology). 
\end{acknowledgments}

\end{document}